\documentclass[aps,twocolumn,showpacs,floats,superscriptaddress, nofootinbib]{revtex4}
\usepackage{epsfig,color}

\def\p{{\bf p}}

\def\8{\infty}
\def\oh{\frac{1}{2}}

\def\undertext#1{\vtop{\hbox{#1}\kern 1pt \hrule}}

\def\dd#1{\frac{d}{d#1}}
\def\dbyd#1#2{\frac{d#1}{d#2}}

\def\p2byp#1#2{\frac{\partial^2#1}{\partial#2^2}}

\def\be{\begin{equation}}
\def\ee{\end{equation}}
\def\bea{\begin{eqnarray} & &}
\def\eea{\end{eqnarray}}

\def\rf#1{(\ref{#1})}

\def\rf#1{(\ref{#1})}

\def\rfs#1{Eq.~\rf{#1}}

\begin{document}


\title{Phase diagram of the disordered Bose-Hubbard model}


\author{V. Gurarie}
\affiliation{Department of Physics, University of Colorado,
Boulder CO 80309, USA}

\author{L. Pollet}
\affiliation{Physics Department, Harvard University, Cambridge MA 02138, USA}
\author{N.V. Prokof'ev}
\affiliation{Department of Physics, University of Massachusetts,
Amherst, MA 01003, USA}
\affiliation{Russian Research Center ``Kurchatov Institute'',
123182 Moscow, Russia}

\author{B.V. Svistunov}
\affiliation{Department of Physics, University of Massachusetts,
Amherst, MA 01003, USA}
\affiliation{Russian Research Center ``Kurchatov Institute'',
123182 Moscow, Russia}

\author{M. Troyer}
\affiliation{Theoretische Physik, ETH Zurich, 8093 Zurich, Switzerland}


\date{\today}

\begin{abstract}
We establish the phase diagram of the disordered three-dimensional Bose-Hubbard
model at unity filling which has been controversial for many years. The theorem of inclusions, proven in Ref.~\cite{Pollet2009}, states that the Bose glass phase always intervenes between the Mott insulating and superfluid phases. Here, we note that assumptions on which the theorem is based exclude phase transitions between gapped (Mott insulator) and gapless phases (Bose glass). The apparent paradox is resolved through a unique mechanism: such transitions have to be
of the Griffiths type when the vanishing of the gap at the critical point is due to a zero
concentration of rare regions where extreme fluctuations of disorder mimic a {\it regular}
gapless system.
An exactly solvable random transverse field Ising model in one dimension is used to
illustrate the point. A highly non-trivial overall shape of the phase diagram is
revealed with the worm algorithm. The phase diagram features a long superfluid finger at
strong disorder and on-site interaction.
Moreover, bosonic superfluidity is extremely robust against disorder in a broad range of interaction parameters; it persists in random potentials nearly 50 (!) times larger than the particle half-bandwidth.
Finally, we comment on the feasibility of obtaining this phase diagram
in cold-atom experiments, which work with trapped systems at finite temperature.
\end{abstract}

\pacs{05.30.Jp, 63.50.-x, 03.75.Hh }

\maketitle

\section{Introduction}\label{sec:introduction}

The behavior of interacting bosons subject to static disorder is a fascinating subject whose study started more than 20 years ago~\cite{Giamarchi1988,Weichman1989}.
An important question raised in these papers is whether a direct transition between the gapped Mott insulating (MI) and superfluid (SF) phases is possible in the presence of disorder.
Fisher {\it et al.} \cite{Weichman1989} argued that a direct
transition was unlikely, though not fundamentally impossible.
Since then, the issue was a topic of hot debate with numerous analytical, computational, and
experimental results reaching contradicting conclusions~\cite{Freericks1996, Scalettar1991, Krauth1991, Zhang1992, Singh1992, Makivic1993, Wallin1994, Pazmandi1995, Pai1996, Svistunov1996, Pazmandi1998, Kisker1997, Herbut1997, Trivedi1997, Sen2001, Lee2001, PS04, Wu2008, Bissbort2009, Weichman1, Weichman2}. Curiously, a large number of direct ~\cite{Krauth1991, Makivic1993, Wallin1994, Pai1996, Trivedi1997, Sen2001, Lee2001} and some approximate approaches~\cite{Zhang1992, Singh1992, Pazmandi1995, Pazmandi1998, Bissbort2009} observed this unlikely scenario!

In  Ref.~\cite{Pollet2009} a final verdict was cast by proving analytically that for
any generic bounded disorder a direct transition between a superfluid and a gapped insulating phase is not possible.
Generic disorder is characterized by an arbitrary non-vanishing probability
distribution of disordered fields within the bounds. Careful direct numerical simulations were in line with this prediction:
in the presence of disorder, no matter how small, a Bose glass (BG) phase always intervenes between the superfluid and Mott insulator phases.
The Bose glass phase is an insulator with localized particle states at the chemical potential. Depending on
system parameters these states can best be described either as localized single-particle levels or as
isolated superfluid lakes. While the Bose glass does not allow for phase coherence to extend over the entire  system,
it is characterized by a finite density of states and thus a finite compressibility and
gapless particle and hole excitations. The result of Ref.~\cite{Pollet2009} comes as a simple corollary of the {\it theorem of inclusions},
which states that for any transition in a system with generic disorder one can always find rare regions of the competing phase on either side of the transition line, provided the position of the line depends on the disorder distribution function. However, there is a certain subtlety, if not a contradiction:
The theorem seems to exclude {\it any} transition between gapless and gapped phases in disordered systems,
and the question arises of how to reconcile the theorem with the phase transition between the gapped Mott insulator and the gapless Bose glass phase.

Previously it was conjectured \cite{Weichman1989,Weichman1,Weichman2}, but never proven rigorously, that the Mott insulator -- Bose glass transition occurs when the bound $\Delta$ on disorder in the local chemical potential equals $E_{g/2}$.  Here $E_{g/2}=\min(E_p,E_h)$ is the smaller of the particle ($E_p$)  and hole  ($E_h$)  excitation gaps in the ideal Mott insulator  (assuming that one works in the grand-canonical ensemble)
\footnote{In Ref.~\cite{Weichman1}, the convexity of free energy as a function of $\varepsilon_i$---the crucial assumption in that paper--- is a conjecture that might hold for the Bose-Hubbard model,
but is incorrect in general, as shown by several counter examples.}.
If we denote by $\mu_+$ and $\mu_-$ the chemical potential thresholds for doping the Mott insulator with particles and holes respectively, then $E_p=\mu_+ - \mu$, $E_h=\mu - \mu_-$. The gap for creating a
particle-hole excitation (the MI gap), $E_g=E_p +E_h = \mu_+ - \mu_-$, is independent of the global chemical potential $\mu$. At zero temperature, the chemical potential of the Mott insulator state with
integer filling factor can be anywhere between the two thresholds leading to an ambiguity in
the value of $E_{g/2}$. The ambiguity is absent in the canonical ensemble, where particle and hole excitations can be created only in pairs, to preserve the total number of particles.
The grand-canonical counterpart of the canonical situation corresponds to the chemical potential being kept  in
the middle of the gap, $\mu = (\mu_+ - \mu_-)/2$,
in which case $E_p=E_h=E_{g/2}=E_g/2$. Therefore, below we always assume this choice of $\mu$.

The above-mentioned $\Delta_c=E_{g/2}$ conjecture is based on the assumption that the state remains gapped for $\Delta < E_{g/2}$.
For $\Delta > E_{g/2}$ the state can be shown to be gapless, because rare statistical
fluctuations guarantee the existence of arbitrarily large homogeneous regions with disorder mimicking chemical potential shifts exceeding particle or hole gaps. In other words
the conjecture was that the transition is of the Griffiths type. An alternative scenario would
claim that the transition point happens at smaller values of $\Delta$ due to subtle interplay
between disorder and interactions.

In this paper, we show that the theorem of inclusions forces one to conclude that the Griffiths-type
scenario is the only one possible for the gapped-to-gapless transitions. That is, the vanishing of the gap at the critical point is exclusively due to a zero concentration of  rare regions in which extreme
fluctuations of disorder reproduce a regular gapless system.
In the vicinity of the critical point, the gapless phase must necessarily be ``glassy", because it consists of large gapless (in our case superfluid) domains embedded in a gapped state.
The absence of phase coherence between domains is caused by their diverging distance between at the critical line.
To illustrate these general conclusions, we consider the exactly solvable random transverse
field Ising model in one dimension.

\begin{figure}
\includegraphics[scale=0.35, angle=-90]{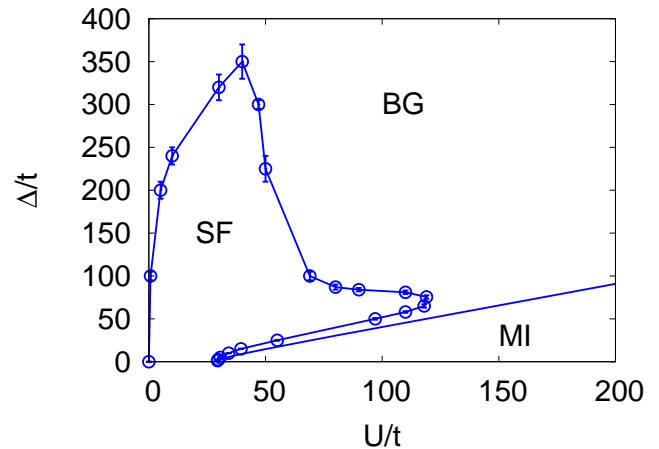}
\caption{Phase diagram of the disordered three dimensional Bose-Hubbard model at unity filling. In the absence of disorder, the system undergoes a quantum phase transition between  SF and  MI phases. The presence of disorder allows for a compressible, insulating BG phase, which always intervenes between the MI and SF phases because of the theorem of inclusions~\cite{Pollet2009}. The transition between MI and BG is of the Griffiths type, as an exception implied by the theorem. At $U/t \to 0$, the SF--BG transition line has an infinite slope \cite{PN09}. }
\label{fig:phasediagram}
\end{figure}

Though the topology of the phase diagram for the Bose-Hubbard model is fixed by
theorems, it is both interesting and important to determine transition lines and properties of phases
numerically. In particular, this is necessary for  revealing
potential difficulties in observing and identifying the phases. To this end, we have
calculated the full phase diagram of the disordered three-dimensional Bose-Hubbard model, shown in Fig.~\ref{fig:phasediagram}, by quantum Monte Carlo simulations based on the worm algorithm
\cite{worm,worm_lode}. This phase diagrams shows a few remarkable features: an infinite slope of the superfluid -- Bose glass line $\Delta_c(U)$,
in the weakly interacting gas $ U/t \lesssim 1$, as predicted by the scenario of percolating superfluid lakes developed in Ref.~\cite{PN09}, and an enormous scale for the superfluid -- Bose glass transition, $\Delta/t \sim 300$  at intermediate coupling strength, $1 \lesssim U/t \lesssim 30$.
Here $U$ is the strength of the on-site repulsion between bosons and $t$ is the amplitude of hopping transitions between the nearest neighbor sites (see Fig.~\ref{fig:phasediagram}).
The percolation character  of superfluidity in the vicinity of the superfluid to Bose glass transition, is most likely the reason for the enormous scale. In this range of parameters,  the localized states have a localization length of the order of one lattice spacing, as opposed to the picture of large superfluid lakes of  Ref.~\cite{PN09}. 

The nature of the transitions and small superfluid fraction in the SF phase have profound implications for the experimental observation of the phase diagram. We  focus here on cold-atom experiments, where recent experimental claims are partly in line, partly in contradiction with the phase diagram shown above. We argue that present-day cold-atom experiments face numerous difficulties in obtaining the full phase diagram; for example, the Griffiths type Bose glass -- Mott insulator transition requires macroscopically large system sizes to properly identify the Bose glass phase. We also provide arguments why experiments seem to have missed the superfluid `finger' above the Mott insulator in Fig.~\ref{fig:phasediagram}, though the right scale for the transition between the superfluid phase and the Bose glass phase for very strong disorder has been revealed~\cite{Pasienski2009}.

The paper is organized as follows. In Sec.~\ref{sec:model} we introduce the model and recapitulate the theorem of inclusions. The transition between the Mott insulator and Bose glass phases is discussed in Sec.~\ref{sec:griffiths} and illustrated by the exactly solvable random transverse Ising model in one dimension. We proceed with a discussion of the full phase diagram in Sec.~\ref{sec:fullphasediagram} and results of cold-atom experiments in Sec.~\ref{sec:experiments}. The conclusions are presented in Sec.~\ref{sec:conclusions}.

\section{Model and theorem of inclusions}\label{sec:model}

The disordered Bose-Hubbard model on a simple cubic lattice is defined the Hamiltonian
\be \label{eq:ham} H =-t \sum_{\langle jk \rangle} \hat a^\dagger_j \hat a_k+ \sum_j  \left(  \epsilon_j -\mu \right) \hat n_j + \frac{U}{2} \sum_j \hat n_j (\hat n_j-1) ,
\ee
where $\hat a_j^\dagger$ is the creation operator of a boson on a site $j$; the symbol
$\langle\ldots \rangle$ denotes summation over nearest neighbor pairs of sites; $\hat n_j=\hat a^\dagger_j \hat a_j$ is the boson density operator; and $\epsilon_j$ is the disordered on-site potential. Without loss of generality, we take $\epsilon_j$ to be independent random variables distributed according to the probability density $p(\epsilon/\Delta)$. The probability distribution satisfies the normalization condition $\int_{-1}^{1} du \, p(u)=1$, has zero first moment $\int_{-1}^{1} du \, u p(u)=0$ (otherwise it is absorbed in the definition of $\mu$), and is taken to be  bounded,
that is $p(u)=0$ if $\vert u \vert > 1$. Formally, the disorder bound $\Delta$ and the disorder distribution dispersion $\delta$ are independent parameters. For the most common
choice of the uniform distribution $p(u)={\rm const}$ (used in our numerical simulations as well),
we have $\delta^2=\Delta^2/3$. A complete characterization of disorder is based on infinite number of parameters fixing the shape
of $p(u)$. One may also add parameters which control correlations between potentials on
different lattice sites, etc. Collectively, we denote all these parameters by $\xi$ and
identify them with the definition of a particular model of disorder.

\begin{figure}
\includegraphics[scale=0.35]{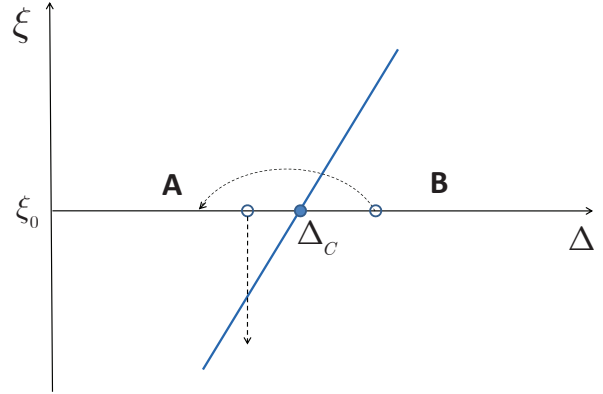}
\caption{ A sketch of the generic phase transition line between some phases A and
B in the plane of disorder distribution parameters $\Delta$ and $\xi$, where $\Delta$ is the
bound and $\xi$ is one of the infinite number of parameters characterizing the
disorder distribution function, e.g. dispersion, and its spatial correlations.
Dashed lines with arrows originate from points which determine disorder properties in the
macroscopic (thermodynamic limit) system and end at points which characterize disorder
parameters in an arbitrarily large, but finite, domain as a result of a rare statistical
fluctuation in the same system.}
\label{fig:2}
\end{figure}

Suppose now that a disordered system, described by the Hamiltonian similar to \rfs{eq:ham},
undergoes a transition from phase A to phase B --- let us for the moment not specify the nature of phases --- as the disorder bound $\Delta$ increases, and that the transition happens at a critical point $\Delta_c$, see Fig.~\ref{fig:2}. One obviously expects that $\Delta_c$ depends on the
disorder strength $\delta$, correlations between the sites, etc. For example, if correlations
are long ranged and $p(u)$ is very close to a $\delta$-distribution,
we hardly have any
disorder at all no matter what the bound is, while for the uniform uncorrelated distribution
disorder can radically change system properties for large $\Delta$.
Let us define the notion of the generic A--B transition as a transition with some
$\Delta_c (\xi )$ dependence.
Figure \ref{fig:2} then proves that close to the transition, if $\Delta>\Delta_c$, there exist domains in phase B which locally look like phase A. Indeed, in the system described by \rfs{eq:ham} at $\Delta>\Delta_c$ one can always find statistically rare domains where
disorder realization is such that, within that domain, $\epsilon_i$ represent
a typical realization of another disorder distribution with a bound $< \Delta_c$, see the dashed line going from B to A in Fig.~\ref{fig:2}.
Locally these domains are in the phase A. The probability of observing such domains
decreases exponentially with their size.

What is more important is that close enough to the transition when $\Delta< \Delta_c$, there exist domains in phase A which locally are in phase B. At first glance, this is hard to justify, because the argument of the previous paragraph can not be used as $\Delta$ has to be less than $\Delta_c$ everywhere. However, if we think in terms of {\it all} possible models of
generic disorder we recognize that the actual value of $\Delta_c$ depends on the details of the distribution function $p(u)$. This implies that it is always possible to choose $\xi$  such that $\Delta_c(\xi)< \Delta$, and thus there are going to be domains in phase A, albeit exponentially rare as they get larger, where local disorder is indistinguishable from a typical realization of disorder with the distribution $p_\xi(u)$ and $\Delta_c(\xi) < \Delta$, see the dashed line going from phase A to phase B in Fig.~\ref{fig:2}. These domains will contain phase B.

The above argument shows that it is not possible for phase A to be gapped if phase B is gapless. Indeed, the B-domains of arbitrarily large size within phase A guarantee that phase A is also gapless. As a consequence, no direct transition between the gapped Mott insulator phase and the gapless superfluid phase is possible.

\section{Griffiths transitions}\label{sec:griffiths}

\subsection{An exception implied by the rule}

The theorem of inclusions rests on the dependence of the critical point on disorder properties
such as its dispersion, correlations, etc. Still one expects the gapful-to-gapless MI--BG
transition to exist,  in apparent contradiction with the theorem! The paradox is resolved
by considering the only remaining possibility, namely, that the transition point $\Delta_c$ does
not depend on $\xi$! In this case one cannot use arguments of the previous section to prove
that in the vicinity of the transition point one can find arbitrary large domains of gapless
phase B (we identify B with the Bose glass) inside the A (identified with the Mott insulator) \footnote{Given an infinite number of continuous parameters determining disorder properties the probability that any particular model of disorder $\xi_0$ is a minimum is zero.}.

The transition which depends only on the bound $\Delta$ cannot be linked to any local physics,
because as the dispersion $\delta$ goes to zero the system becomes indistinguishable from a pure one
on larger and larger scales. This forces one to conclude that the transition mechanism itself is
necessarily based on rare statistical fluctuations which explore the possibility of reaching the
disorder bound at all sites on larger and larger scales.
Suppose that a gapped phase can be rendered gapless by applying a
regular external field $H$. For the Mott insulator such a field is a global chemical potential
shift $\delta \mu$; whenever $\mu+\delta \mu$ is above $\mu_+$ or below $\mu_-$ the system
is doped with particles or holes and enters the superfluid state. The pathological
insensitivity of the critical value $\Delta_c$ on $\xi$ is natural for this scenario
of rare regions in which the disorder fluctuation is reproducing a {\it regular pure} system in an
external field. When the disorder bound allows one to reach the critical value
of the field, a transition occurs. We recognize that this mechanism is nothing but
the conjectured Griffiths type  MI--BG transition when the vanishing of the gap at the critical point
is  due to an infinitesimal concentration of rare regions in which the fluctuation of disorder mimics a {\it homogeneous} chemical potential shift \cite{Weichman1989}. In the general case it can be any regular external field whose amplitude scales with $\Delta$.

We thus conclude that that gapless-to-gapful transitions in disordered systems are possible
if, and only if, they are of the Griffiths type and the transition line is fully determined by
the properties of a pure system. In this case the disorder bound protects the gapped phase A (the Mott insulator) from having rare regions of phase B embedded in it. At the same time, when $\Delta$ is only slightly larger than $\Delta_c$, then phase B appears to be identical to phase A locally except that it has rare, well-separated regions containing a gapless {\it pure} system.
This means B cannot be superfluid, i.e. it is a glassy state.

\subsection{Illustration} \label{sec:illustration}

To illustrate the arguments presented above, we consider a disordered one-dimensional quantum
model that shares some features with \rfs{eq:ham}, but is exactly solvable.
It is closely related to the the two-dimensional classical Ising model with bonds whose strength depends randomly on their position in one spacial direction while being independent of the position in the second spatial direction. This model was first solved in Ref.~\cite{Mccoy1968}. It is characterized by a high temperature gapped paramagnetic phase, low temperature gapped ferromagnetic phase, and the intermediate Griffiths phase \cite{Griffiths1969}.
The Griffiths phase is akin to the Bose glass phase in model \rfs{eq:ham}, while the gapped phases are similar to the Mott insulator.
The nature of the transition between these phases can thus be clarified with the help of the exact solution (see also the discussion in Ref.~\cite{FisherD1995}).

The model we consider here is a continuum limit in the second spatial direction, which we interpret as time. Then it is equivalent to the random transverse field one-dimensional Ising model, which we can write as 
\be \label{eq:tfim} H = \sum_j \left( t \, \sigma^x_j \sigma^x_{j+1} + h_j \sigma^z_j \right).
\label{Isingmodel}
\ee
Here $\sigma^x_j$ and $\sigma^z_j$ are Pauli matrices, acting on a $j$-th site of a linear chain, while $h_j$ are random independent variables. The probability distribution $p(h)$
is taken to be uniform on the $h \in \left[ h_0-\Delta, h_0+\Delta \right]$ interval, where $h_0$ and $\Delta$ are positive parameters such that $h_0> \Delta$. In principle, one could also add disorder to the Ising coupling $t$, however, this is not needed for the purpose of our illustration here.

Model (\ref{Isingmodel}) is solved exactly by the Jordan-Wigner transformation~\cite{Lieb1964}, which became standard  for these types of problems. Let us briefly review this method.
With the notations
$\sigma^+=(\sigma^x+i\sigma^y)/2$, $\sigma^-=(\sigma^x-i \sigma^y)/2$ we introduce the Jordan-Wigner fermions
\be \hat a_j =  \sigma^-_j \, e^{i \pi \sum_{k<j}\sigma^+_k \sigma^-_k}, \ \hat a_j^\dagger =  \sigma^+_j \, e^{i \pi \sum_{k<j}\sigma^+_k \sigma^-_k}.
\ee
They satisfy the usual fermionic anticommutation relations
\be \left\{ a^\dagger_j , \hat a_k \right\}_+ = \delta_{jk}, \  \left\{ \hat a_j , \hat a_k \right\}_+ = 0.
\ee
In terms of these, the transverse field Ising model becomes
\be
H = \sum_j h_j \left( \hat a^\dagger_j \hat a_j - \hat a_j \hat a^\dagger_j \right)+ t \sum_j  \left( \hat a^\dagger_j - \hat a_j \right) \left(\hat a^\dagger_{j+1}+\hat a_{j+1} \right).
\ee
This Hamiltonian has the standard Bogoliubov form familiar from the theory of superconductivity
and can be rewritten as
\be
H = \sum_{jk} \left( \matrix{ \hat a_j^\dagger & \hat a_j  }\right)  {\cal H}_{jk}
\left( \matrix {\hat a_k \cr \hat a^\dagger_k }\right),
\ee
where
\be
{\cal H}_{jk} = \oh \left( \matrix {{\cal D}_{jk} + {\cal D}^T_{jk}  & {\cal D}_{jk} - {\cal D}^T_{jk}\cr {\cal D}^T_{jk} - {\cal D}_{jk} & - {\cal D}_{jk} - {\cal D}^T_{jk}} \right).
\ee
and ${\cal D}$ is a matrix defined as
\be \label{eq:defd} {\cal D}_{jk} = h_j \delta_{jk} +t \, \delta_{j,k-1}.
\ee
The problem now reduces to diagonalizing the real symmetric matrix ${\cal H}$. To do so, it is convenient to perform first a unitary (actually, in this case, orthogonal)
transformation defined as
\be U = \frac 1 {\sqrt 2} \left( \matrix{ 1 & 1 \cr -1 & 1 } \right), \tilde {\cal H}= U^T {\cal H} U \;.
\ee
This gives
\be \label{eq:zi}  \tilde {\cal H} =\left( \matrix{ 0 & {\cal D} \cr {\cal D}^T &  0 } \right).
\ee
We recognize in ${\cal H}$ a random one-dimensional Hamiltonian in the BDI symmetry class, according to the classification scheme of Ref.~\cite{Zirnbauer1996}. This means that $ \tilde{\cal H}$ is real and that there exists a matrix $\Sigma^3$, in case of \rfs{eq:zi} given by
\be \Sigma^3 = \left( \matrix { 1 & 0 \cr 0 & - 1} \right),
\ee such that \be \label{eq:symmetry} \Sigma^3  \tilde {\cal H} \Sigma^3 = -  \tilde {\cal H}.\ee  The arguments of Ref.~\cite{Zirnbauer1996}
relate the peculiar properties of the spectrum of the Hamiltonian \rf{eq:zi} which are discussed below to the existence of the symmetry \rf{eq:symmetry}.

The problem defined by the Hamiltonian \rfs{eq:zi} together with \rfs{eq:defd} was solved exactly over thirty years ago in Ref.~\cite{Eggarter1978}
by the transfer matrix techniques, now standard in one dimensional disordered systems.
In principle we could use this solution to extract all the information we need about the random transverse field Ising model \rfs{eq:tfim} and its phase transitions. Yet the solution presented in Ref.~\cite{Eggarter1978} is still relatively involved. To illustrate the main features of the phase transitions, we can
go to the continuum limit of \rfs{eq:zi}. In the continuum, the corresponding problem was solved in Ref.~\cite{Comtet1995}. Their method is very simple and versatile, so we would like to put it to use here.

The continuum limit in the Hamiltonian $\tilde {\cal H}$ occurs close to the center of the band or to the momentum $\pi$, when $h_j$ do not deviate much from some average value $h_0$. More formally, we need to further transform the Hamiltonian by the unitary transformation given by
\be U = \left[ \matrix { (-1)^j \delta_{jk} & 0 \cr 0 & (-1)^j \delta_{jk}} \right],
\ee
which keeps the structure of $\tilde {\cal H}$ intact but with the matrix ${\cal D}$ now given by
\be \label{eq:defd1} {\cal D}_{jk} = h_j \delta_{jk} -t \, \delta_{j,k-1} =-  t \left( \delta_{j,k-1}-\delta_{jk} \right) + \left( h_j - t\right) \delta_{jk}.
\ee
Now it is clear that the continuum limit of our problem is given by the same matrix $\tilde {\cal H}$  of \rfs{eq:zi} with
\be {\cal D} = ta \left[ -  \dd{x} + V(x) \right]. \ee
Here the continuum variable $x$ is taken to be equal to $j$ and  $a$ is the lattice spacing, while
\be V(x) = \frac {h_j - t}{ta}.\ee
 In the continuum, $V(x)$ can be thought of as a spatially random potential.

Now the methods of Ref.~\cite{Comtet1995} (as adopted for this problem in Ref.~\cite{Gurarie2003a}) can be brought to bear on this problem. One result is that the spectrum of $\tilde {\cal H}$ is fully gapped if $V(x)$ is everywhere positive or everywhere negative. Recalling the definition of $V(x)$ via $h_j$ and
properties of the probability distribution $p(h)$, this implies $h_0> t+\Delta$ or $h_0 < t-\Delta$. In other words,  $h_j$ are either all greater than $t$ or all smaller than $t$.

Consider, for example, the case of $V>0$. Suppose one increases $\Delta$ until regions appear where $V(x)<0$ (or $h_j <t$). As soon as they appear, the spectrum of ${\cal H}$ becomes gapless. The density of states for positive energies $E$ can be computed using the following construction. Take all regions where $V(x)<0$. Consider the probability $P(E)$  that
\be \label{eq:prob} \ln E > \int_{x_1}^{x_2} dx \, V(x),
\ee
where the integration goes over one of the continuous  intervals $[x_1,x_2]$ where $V(x)<0$.
Then the density of states is given by
\be \label{eq:dens} \rho(E) \sim \dbyd{P(E)}{E}\;,
\ee
with $\alpha \ge -1$.
We expect this probability to be exponentially small in $\ln E$, or
\be \label{eq:exp} P(E) \sim \exp \left[ \left( \alpha+1 \right) \ln E \right],
\ee where $\alpha$ is some number. Then
\be \label{eq:rho} \rho(E) \sim E^\alpha.
\ee
The derivation of Eqs.~\rf{eq:prob}, \rf{eq:dens} and \rf{eq:rho} is given in Ref.~\cite{Gurarie2003a}, and this completes the exact solution.

We are now in a position to fully describe the transition from a gapped paramagnetic phase with all $h_j>t$ to the gapless Griffiths phase as the disorder strength $\Delta$ is increased.
For $\Delta > h_0-t$, large rare regions appear where $V(x) \sim h_j - t$ is negative. We can use \rfs{eq:prob} to calculate $\alpha$, whose precise value depends on the probability distribution $p(h)$ but which in general is equal to some large number decreasing as $\Delta$ is increased past $h_0-t$. Thus the low energy states which appear as $\Delta$ is increased above the threshold $h_0-t$ will be suppressed by a power law, according to \rfs{eq:rho}. The Griffiths phase we obtained in this way is characterized by gapless excitations whose density is suppressed at low energy. Sometimes such a phase is referred to as a phase with a pseudogap (similar to a Mott glass phase arising in
systems with exact particle-hole symmetry and off-diagonal disorder \cite{Giamarchi2001,PS04}).


We observe that the transition from the gapped paramagnetic phase to the gapless (but glassy) Griffiths phase proceeds exactly via the route described in this paper. When $\Delta<h_0-t$, no disorder, no matter what the details of its distribution are, can create gapless states. The transition to the Griffiths phase occurs when disorder is just strong enough to create regions
where gapless excitations can reside, because in this region an effective field $h=h_0-\Delta$
can be made arbitrarily close to the critical value $h_c=t$. We note that the difference between the Bose-Hubbard model \rfs{eq:ham} and the random transverse field Ising model \rfs{eq:tfim} lies in the fact that \rfs{eq:tfim}, even in the fully clean (no disorder) regime, does not have a truly gapless phase, such as the superfluid in the Bose-Hubbard model. Yet the fact that \rfs{eq:tfim} has a critical point in the absence of disorder is sufficient to create a glassy Griffiths phase with gapless excitations described by the power-law density of states.

Identical arguments describe the transition from the gapped ferromagnetic phase to the Griffiths phase if $h_0 <t$ as disorder strength $\Delta$ is increased past $t-h_0$.

\section{Global phase diagram}\label{sec:fullphasediagram}

In view of ongoing experimental activity to study the physics of interacting disordered bosons
in optical lattice, and to connect the limit of strong interactions where disorder competes with
the physics of Mott insulators with the physics of localization in weakly interacting systems, we
performed first-principles quantum Monte Carlo simulations of the model (\ref{eq:ham}). The results at unity filling  are presented in
Fig.~\ref{fig:phasediagram}. As predicted by the theorem of inclusions, the superfluid and Mott insulator phases
are always separated by the Bose glass phase at any $\Delta > 0$.

\begin{figure}
\includegraphics[scale=0.45]{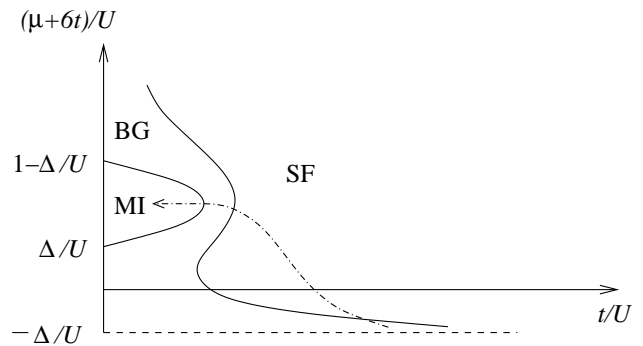}
\caption{The phase diagram $(\mu+6t)/U$ vs $t/U$  of the Bose-Hubbard model, adapted from Ref.~\cite{Weichman1989}. A dashed-dotted line represents
the process whereby  the interaction strength $U$ is increased at a fixed disorder $\Delta$ in a system with unity filling factor, in the regime of the reentrant SF--BG transition.}
\label{fig:global_p}
\end{figure}

An interesting feature of the phase diagram from Fig.~\ref{fig:phasediagram} is the reentrant nature of the Bose glass -- superfluid transition if the interaction
strength is increased at fixed disorder (and as long as disorder is not too strong to suppress the SF phase completely), confirming previous studies~\cite{Krauth1991, Krauth1991_EPL}. Fig.~\ref{fig:global_p} shows this
reentrant behavior on the more familiar $(\mu+6t)/U$ vs $t/U$ phase diagram. The dashed-dotted line represents a line of unit filling, $\langle n \rangle = 1$, as the interaction strength is increased for fixed disorder strength.

If disorder strength is increased, the Bose glass -- superfluid boundary line moves off to the right, and the entire line representing the Bose-Hubbard model at varying $U$ may end up inside the Bose glass phase.

Somewhat surprisingly, we find that
the superfluid phase extends to the region of very strong interactions and disorder where
both $U$ and $\Delta$ are about two orders of magnitude larger then the hopping amplitude.
This narrow finger-like region has fragile superfluid properties: being surrounded by the
insulating
state it has to have low transition temperatures to the normal state and
small superfluid density at $T=0$. Indeed, both quantities go to zero at the phase boundary.
The superfluid transition temperature at $\Delta/t=65$ and $U/t=60$
in the middle of the finger base is as low as $T_c/t\approx 0.37(5)$.
Correspondingly, weak coherence properties (small condensate fraction) are expected in the
finger region. Moreover, they can be observed only on sufficiently large scale, because
the correlation length diverges along the boundary.

In another region of the phase diagram in Fig.~\ref{fig:phasediagram},  for $U \to 0$, the data clearly indicates an infinite derivative of the
$\Delta_c (U)$ curve, in line with the prediction \be \label{eq:PN} \Delta_c \propto U^{1/4} \ee
of Ref.~\cite{PN09}. The results of Ref.~\cite{PN09} were based on Gaussian random disorder, 
as opposed to a bound disorder distribution discussed in this paper.
However, we find that nevertheless their arguments remain qualitatively correct in our case too.
Indeed, the nature of the Bose glass -- superfluid transition at weak interaction strength when $U \ll t$, $\Delta \ll t$, as discussed in Ref.~\cite{PN09}, is based on percolation between localized states with 
energies $E << \Delta$. In this energy range, for states with large localization length, 
the Gaussian character of disorder fluctuations is guaranteed by the central limit theorem.

Quantitatively, we find that extremely large disorder is necessary
to localize bosons even when interactions are relatively weak ($U/t=1$ in terms of the
lattice model parameters corresponds to the gas parameter $na_s^3 \sim 10^{-4}$ for the continuous weakly interacting gas with the s-wave scattering length $a_s$).

At moderate interaction strength and $\Delta \gg t$ we expect that the transition between the
superfluid and the Bose glass phase is still driven by percolation.
Because of the imposed commensurability the mechanism differs from the conventional Anderson localization argument which would predict a critical disorder of the order of the bandwidth.
For strong disorder all single-particle states are localized with the localization length close to unity
(in terms of the lattice constant) \cite{Bulka1987}. The local (site) Hamiltonian
\begin{equation}
H_{\rm loc} = (\epsilon - \mu)n + \frac{U}{2} n(n-1).
\end{equation}
can be used to determine the site occupation number as
\begin{equation}
n = \frac{U/2  + \mu - \epsilon}{U},
\end{equation}
which is valid if $n \ge 0$ or $ \epsilon < U/2 + \mu$. Otherwise, $n=0$. As long as $\Delta \gg U$ the density $n$ can be considered as a continuous function of $\mu$ and there is no need to take into account that $n$ can only be integer. The average density is now
\begin{equation}
\langle n \rangle  = \frac{1}{2\Delta} \int_{-\Delta}^{U/2+\mu} \frac{ U/2 + \mu - \epsilon}{U} d\epsilon .
\end{equation}
Setting $\langle n \rangle$ equal to unity leads to
\begin{equation}
\mu = -U/2 - \Delta + 2\sqrt{U\Delta}.
\end{equation}
A site will be occupied if its disorder lies within the $(-\Delta, U/2 + \mu) =(-\Delta, -\Delta + 2\sqrt{U \Delta} )$ interval. The corresponding probability is $\sqrt{U/\Delta}$.
If we assume that superfluidity requires that occupied sites form a percolating cluster, then
for a simple cubic 3D lattice with the percolation threshold $p_c \approx 0.31$ \cite{Isichenko1992} we find
the transition line at
\begin{equation}
\frac{U}{\Delta} \gtrsim \frac{1}{10}.
\end{equation}
This estimate is in good quantitative agreement with the Monte Carlo results shown in Fig.~\ref{fig:phasediagram} for intermediate coupling before the Mott physics
becomes important at $U/t > U_c/t = 29.34(2)$.

In turn, the assumption made above relies on the fact that moving a boson from one occupied site to another requires energy of the order of $U$, which is independent of disorder, while moving it to an empty site requires a much larger energy of the order of $\Delta$.
We note in passing that while the percolation scenario drives the system towards the Bose glass -- superfluid transition and thus defines, with a certain accuracy, the position of the critical line, the criticality of the transition is most likely to be universal everywhere on the phase diagram.

\section{Implications for cold-atom experiments}\label{sec:experiments}

\begin{figure}
\includegraphics[scale=0.35, angle=-90]{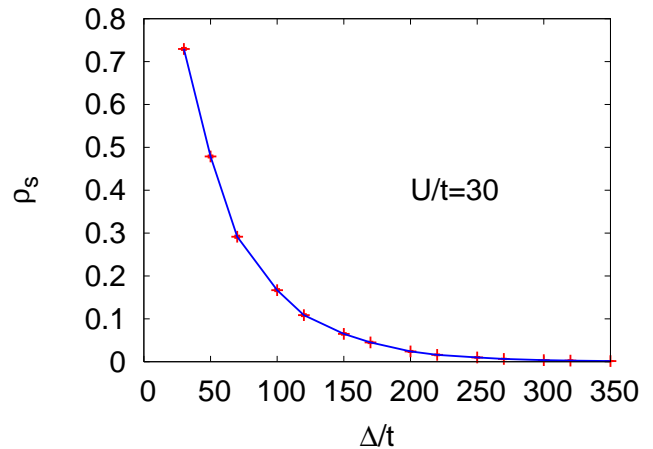}
\caption{Superfluid density as a function of disorder strength $\Delta$ at fixed interaction strength $U/t=30$ for a system size $L=8 \times 8 \times 8$ and inverse temperature $\beta t = 10$. The low value of the superfluid density shows the fragility of the superfluid finger in the large portion of the phase diagram Fig.~\ref{fig:phasediagram}. }
\label{fig:rho_s}
\end{figure}

\begin{figure}
\includegraphics[scale=0.35, angle=-90]{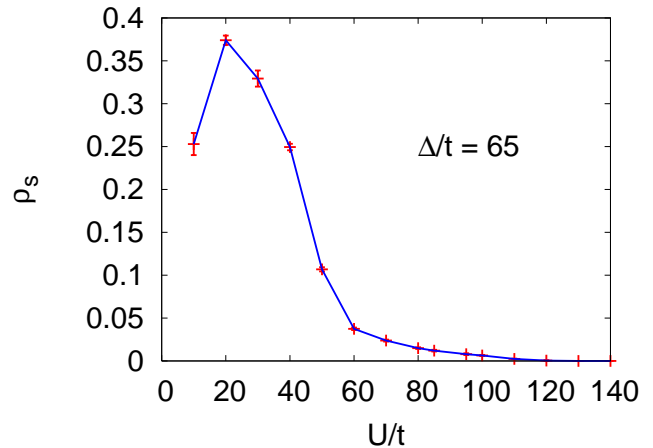} 
\caption{Superfluid density as a function of interaction strength $U$ at fixed disorder strength $\Delta/t=65$ for a system size $L=8 \times 8 \times 8$ and inverse temperature $\beta t = 10$.
The low value of the superfluid density is indicative of weak coherence properties
and low transition temperature to the normal state in the `finger' region.}
\label{fig:rho_s2}
\end{figure}

Recently, experiments with ultracold gases have addressed the disordered three-dimensional Bose-Hubbard model~\cite{White2009,  Pasienski2009}. The random potential is generated using a fine-grained optical speckle field with correlation length comparable in size to the lattice spacing, but the disorder realization is usually kept fixed. The system is probed by looking at interference images giving access to the condensate fraction, $n_0$,
provided the time-of-flight duration is sufficiently long and $n_0$ is large enough to be resolved in a trapped system. Transport properties are obtained by measuring the motion of the centre-of-mass of the atomic cloud immediately after an applied impulse~\cite{McKay2008}.

There are numerous considerations one has to keep in mind when trying to compare
any experimental data to theoretical predictions for the homogeneous thermodynamic system.
The optical speckles not only introduce diagonal site-disorder, but also effect the on-site repulsion strength and the hopping amplitude. In addition, there is a parabolic confinement trap rendering the system mesoscopic and inhomogeneous. This means that there is
often a mixture of phases in the trap, such as the wedding cake structure where commensurate Mott domains are separated by liquid regions. Finally, experiments are done at low, but finite temperature. All of this complicates a direct comparison with the theory.
It is however believed that the experiments can capture the phases and the transitions to some degree.

The lack of a genuine compressibility measurement and a direct measurement of the gap make it difficult for current experiments to distinguish between the Mott insulator and Bose glass phases (they only distinguish between superfluid and insulating phases~\cite{Pasienski2009}). Moreover,
the nature of the Griffiths transition prevents any experiment from direct observation of
the transition line, because this would require astronomically large system sizes.
As discussed above, on short scales the Bose glass phase does appear identical to the Mott insulator phase.
Since differentiating between the Mott insulator and Bose glass phases in the neighborhood of the $U_c/t$ point
is not possible experimentally, we will discuss here only the superfluid -- Bose glass transition.

Experiments find that disorder can induce a superfluid-to-insulator transition, but they see
no evidence for a disorder-generated insulator-to-superfluid transition, in apparent contradiction with Fig.~\ref{fig:phasediagram}. At this point we recall, that superfluidity in the finger
region is easily destroyed even by small finite temperature, because even at the base
of the finger at $\Delta/t=65$ and $U/t=60$ the transition temperature is only $T_c/t\approx 0.40(5)$ -- such low temperatures were never reported in the literature for the
Bose-Hubbard model in the strongly correlated regime. Furthermore coherence is weak even for $T < T_c$, see Fig.~\ref{fig:rho_s2}.
It is likely that both effects are important in understanding why
this region will be missed in the time of flight image.

On the positive side, experiments do find the superfluid -- Bose glass transition for $U/t= 25$ and
$\Delta/U=10$ or $\Delta/t = 250$. From our single site localization argument in combination with percolation it is clear that there is no fundamental problem in observing this transition experimentally at sufficiently low temperature, because it is dominated by the
short-range physics (the size of the superfluid region shrinks at finite temperature).
This finding gives the right order of magnitude answer when compared to the phase diagram shown in Fig.~\ref{fig:phasediagram}. We note that a precise determination of the transition point requires sufficiently big samples that are uniform in the middle (the Monte Carlo results had to be extrapolated to the thermodynamic limit since our answers drifted about ten percent for the lowest system sizes studied), and needs an accurate thermometry to study temperature effects.
Our Monte Carlo results for the superfluid density as a function of disorder at fixed
$U/t=30$, see Fig.~\ref{fig:rho_s}
indicate that $n_s$ is severely depleted at large disorder, and thus
transition temperatures in this region are small.

Some of the difficulties discussed here are specific to the Mott physics at integer
filling factor. Away from commensurability, cold atom experiments can probably be
successful in discerning the insulating glassy phase from the superfluid one.

\section{Conclusions}\label{sec:conclusions}

Summarizing, we have shown that Griffiths-type transitions form a unique exception to the theorem of inclusions. This  immediately implies that all the gapful-to-gapless phase transitions in disordered systems are of the Griffiths type,
and, correspondingly, close enough to the critical point, the structure of the gapless phase is what can generically be referred to as Griffiths glass:
The system of distinct gapless domains containing a regular gapless system embedded
into the gapped phase. This, in particular, proves the Griffiths nature of the Mott-insulator--to--Bose-glass phase transition. With the local shift of the
chemical potential being the relevant field closing the MI gap, the critical line is
given by the condition $2\Delta = E_g$, where $\Delta$ is the bound of disorder and
$E_g$ is the particle-hole gap in the pure system.
We have also considered a particular example of exactly solvable random transverse
field Ising model which perfectly agrees with the established general picture.

The full phase diagram has been presented in Fig.~\ref{fig:phasediagram} and we have discussed
the reason behind extraordinary stability of the superfluid phase against disorder and interactions. In combination with the analytical results in one dimension~\cite{Svistunov1996}, numerical results in 2D~\cite{PS04}, and the theorem of inclusions~\cite{Pollet2009}, this study completes a comprehensive
description of the disordered Bose-Hubbard model at zero temperature in all physically
relevant dimensions.

This work  was supported by the National Science Foundation under
grants PHY-0653183 and DMR-0449521 and by the Swiss National Science Foundation.
The authors acknowledge hospitality of the Aspen Center for Physics  where
this work has been initiated. We are grateful to I. Aleiner, J. Chalker, B. De Marco, E. Demler and P. Weichman for stimulating discussions.
Part of the simulations were performed on the Brutus cluster at ETH Zurich.

\bibliography{dyson}

\end{document}